\documentclass{ws-p8-50x6-00}

\begin{document}

\title{Open Heavy Flavour Production at HERA}

\author{J. Kroseberg}

\address{{\rm (on behalf of the H1 and ZEUS collaborations)}\\ 
Physik--Institut der Universit\"at Z\"urich\\ Winterthurer Str. 190,  
CH--8057 Z\"urich, Switzerland\\ 
E-mail: Juergen.Kroseberg@desy.de}

\maketitle

\abstracts{
Selected recent ZEUS and H1 results on open charm and beauty production in high energy $ep$ collisions 
are reviewed. Measurements of differential $D^*$ cross sections, charm meson production ratios and 
semi--leptonic beauty decays are discussed and compared with QCD predictions.
}

\section{Introduction}
Measurements of charm and beauty production at HERA probe proton and 
photon structure as well as the mechanisms of the hard subprocesses underlying 
$ep$ interactions. 
In particular, they are considered an  excellent testing ground 
for perturbative QCD calculations as the heavy quark mass 
provides a hard scale.
Here, heavy quark production is predominantly gluon induced,  
the leading--order (LO) process being {\it boson--gluon fusion}, 
where  a photon emitted by the electron and a gluon coming from 
the proton form a quark-antiquark-pair. The exchange of a quasi--real photon  
({\it photoproduction}, $\gamma p$) dominates over  
large photon virtualities ({\it deep inelastic scattering}, DIS).

Perturbative QCD calculations are available in next--to--leading order (NLO) 
and yield charm and beauty production cross sections one and three orders of
magnitude smaller than the total cross section respectively. Monte Carlo (MC) simulations 
used in present analyses are based on LO matrix elements, which, except for the 
CCFM--based CASCADE program, are combined with 
a DGLAP--like parton evolution and leading--log parton showers.

The experimental results presented below are based on sub--samples 
of the data recorded by the H1 and ZEUS experiments 
between 1996 and 2000 at $ep$ centre--of--mass energies of 300 and 318 GeV, 
with corresponding integrated luminosities ${\cal L}$ ranging from about 10 to 100 pb$^{-1}$.

\section{Open charm production}
Although recently analyses using other channels 
have become available, most of the HERA results on open charm production are
based on the selection of $D^*(2010)$ mesons via the decay chain 
$D^{*}\rightarrow D^0\pi\rightarrow K\pi\pi$. 

\subsection{The charm proton structure function $F^c_2$}
\begin{figure}[h]
\vspace*{1.3cm}
\epsfxsize=15pc
\hspace*{-0.5cm}\epsfbox{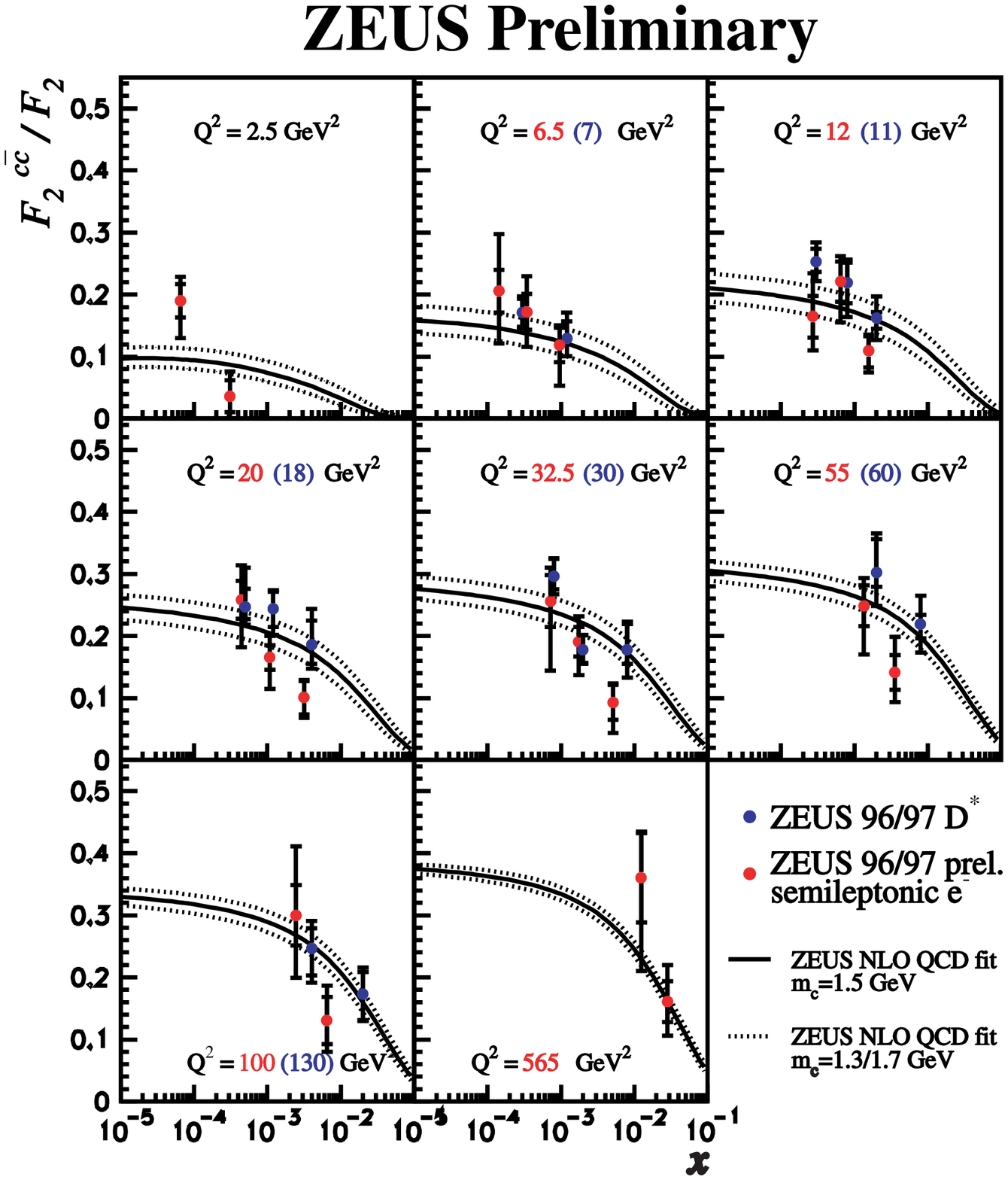} 
\begin{minipage}{5cm}
\vspace*{-5.2cm}
\epsfxsize=19pc 
\hspace*{-1.5cm}\epsfbox{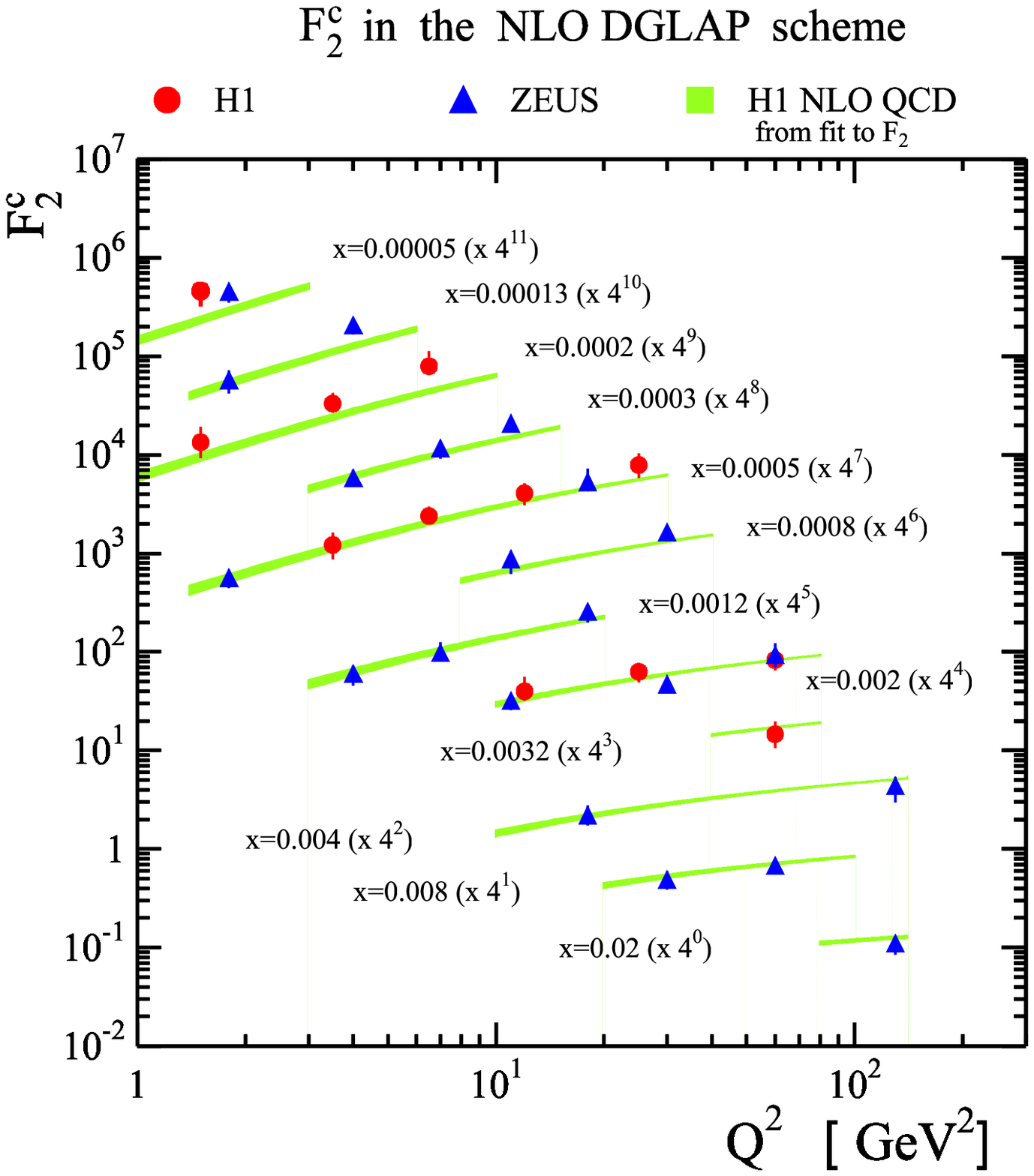} 
\end{minipage} 

\setlength{\unitlength}{1cm}    
\begin{picture}(0.1,0.1)
\put(0.,2.6){(a)}
\put(5.8,2.6){(b)}
\end{picture}
\vspace*{-2.7cm}
\caption{Measured charm contribution to the proton structure function as a function of $x$ and $Q^2$, 
compared with NLO QCD: (a) ratio $F^c_2/F_2$ and  (b) $F^c_2$. 
\label{fig:f2c}}
\vspace*{-0.7cm}
\end{figure}

$D^*$ production in DIS has been studied by both 
H1\cite{h1dstardis} and ZEUS\cite{zeusdstardis}. 
The charm proton structure function $F^c_2$, 
which is defined in analogy to the inclusive structure function $F_2$, 
has been extracted as a function of the Bj\o rken scaling variable $x$ and 
the photon virtuality $Q^2$. 
The charm contribution to DIS is 
significant, the ratio of $F^c_2$ to 
$F_2$  being largest at low $x$ and high $Q^2$, where the gluon component in the proton 
is dominant, cf. figure \ref{fig:f2c}(a). The picture of charm production as 
a gluon driven process is also reflected in 
the scaling violations of $F^c_2$ itself
which are more pronounced than in the inclusive case. As shown in figure 
\ref{fig:f2c}(b), a 
NLO DGLAP QCD fit to inclusive H1 data provides a good description 
of the $F^c_2$ data from both experiments with a tendency, however, to underestimate the rise 
towards higher $Q^2$ at very small $x$.

\subsection{Differential $D^*$ $\gamma p$ cross sections}
\begin{figure}[t]
\epsfxsize=13pc
\hspace*{-0.2cm}\epsfbox{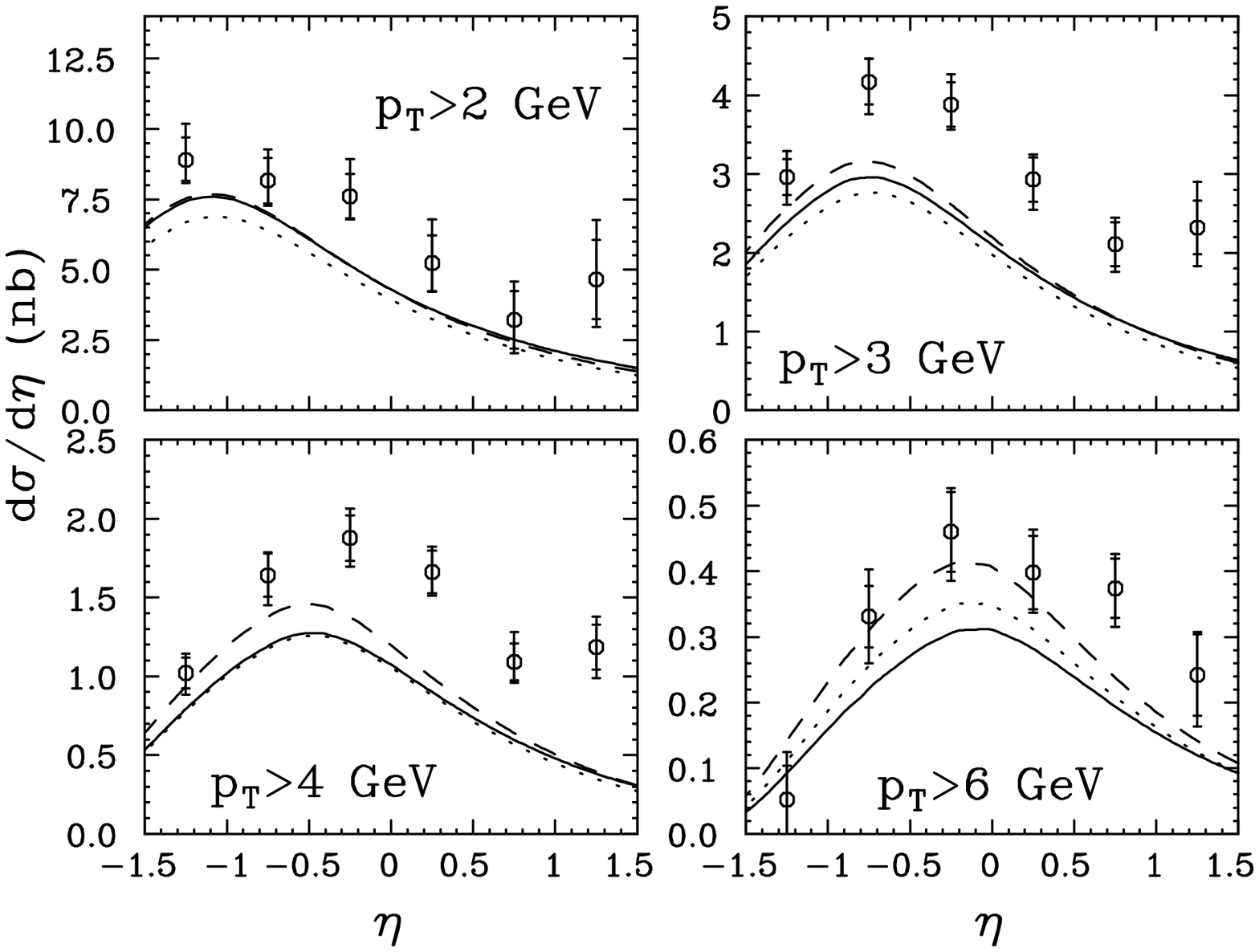} 
\begin{minipage}{5cm} 
\vspace*{-3.5cm}
\epsfxsize=15pc 
\hspace*{0.0cm}\epsfbox{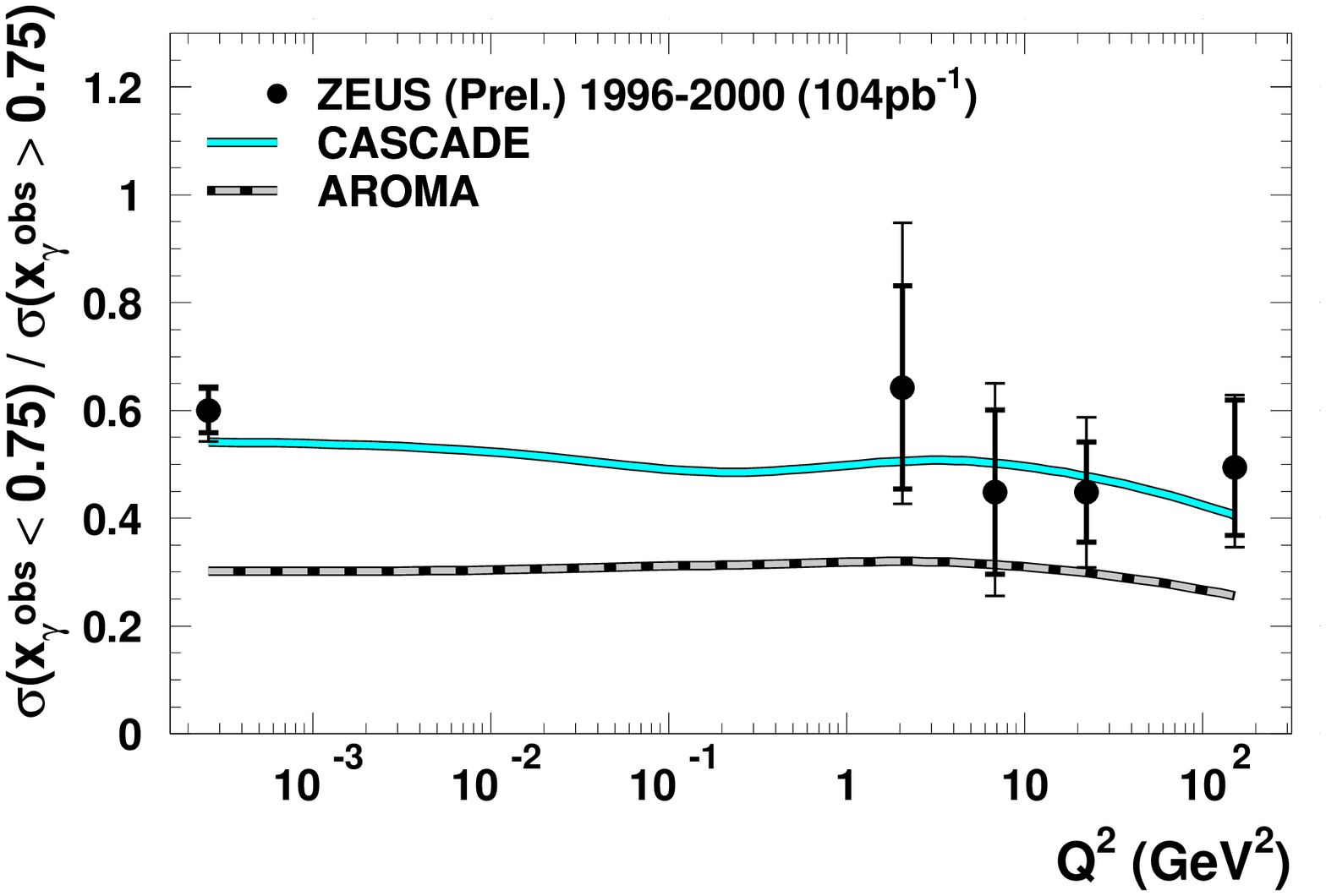} 
\end{minipage}

\setlength{\unitlength}{1cm}    
\begin{picture}(0.1,0.1)
\put(0.,0.3){(a)}
\put(6.2,0.3){(b)}
\end{picture}
\vspace*{-0.3cm}
\caption{(a) $D^*$ cross sections as measured by ZEUS, compared with QCD predictions.
(b) Ratio of resolved to direct $D^*$ production as a function of $Q^2$, compared with LO MC. 
\label{fig:dstargap}}
\vspace*{-0.5cm}
\end{figure}

In photoproduction, double differential cross sections in terms 
of the  $D^*$ transverse momentum and (pseudo)rapidity are derived by H1\cite{h1dstargap}  
and ZEUS\cite{zeusdstargap} and compared to QCD predictions.
Due to differences in observables and visible regions used by the experiments
the data cannot be compared together directly. It has been shown recently, however, that
an FONLL calculation\cite{frixifonll} 
is able to provide an acceptable description of the cross
sections of both experiments. For the H1 results consistency within errors
is found in all measured regions 
while the  ZEUS data hint at a possibly
harder transverse momentum spectrum and an excess in the positive pseudorapidity region, 
cf. figure \ref{fig:dstargap}(a).

\subsection{Resolved photon processes} 
Based on a sample of events with at least two jets and an identified $D^*$ meson, 
ZEUS has studied\cite{zeusdstarjets} the role of resolved photon processes, where, in contrast to direct processes, 
only a fraction $x_\gamma<1$ of the 
photon momentum enters the hard interaction. Experimentally, 
the fraction $x^{obs}_\gamma$ of the incoming  electron momentum going into the production of the two highest-$E_t$ 
jets is used,
which, however, is sensitive to any radiation not collinear to the heavy quarks.
Defining a cut at $x^{obs}_\gamma=0.75$,  
the ratio of resolved to direct processes in $D^*$ photoproduction  is found to be about $60\%$.   
A non--perturbative interpretation of this result in terms of a charm--containing hadronic photon 
structure, however, is compromised by the fact that the measured ratio is 
independent of $Q^2$ within errors and that the CASCADE MC
gives a good description without an explicit resolved photon component, see figure \ref{fig:dstargap}(b).

\subsection{$D$ Meson production ratios}
ZEUS reports results on charm meson production ratios.
Fragmentation fractions for excited meson states have been determined\cite{zeusfragexcite}  
and the relative strangeness production rate has been studied\cite{zeusfragstrange} as well as 
the ratio of vector meson to pseudoscalar meson production.\cite{zeusfragvect}
The results can directly be compared to measurements at $e^+e^-$ colliders 
and good agreement is found in all cases. 
In this sense charm fragmentation appears universal.

\begin{figure}[b]
\vspace*{-1.5cm}
\epsfxsize=11pc
\begin{minipage}{6cm} 
\vspace*{0.3cm}
\hspace*{0.5cm}\epsfbox{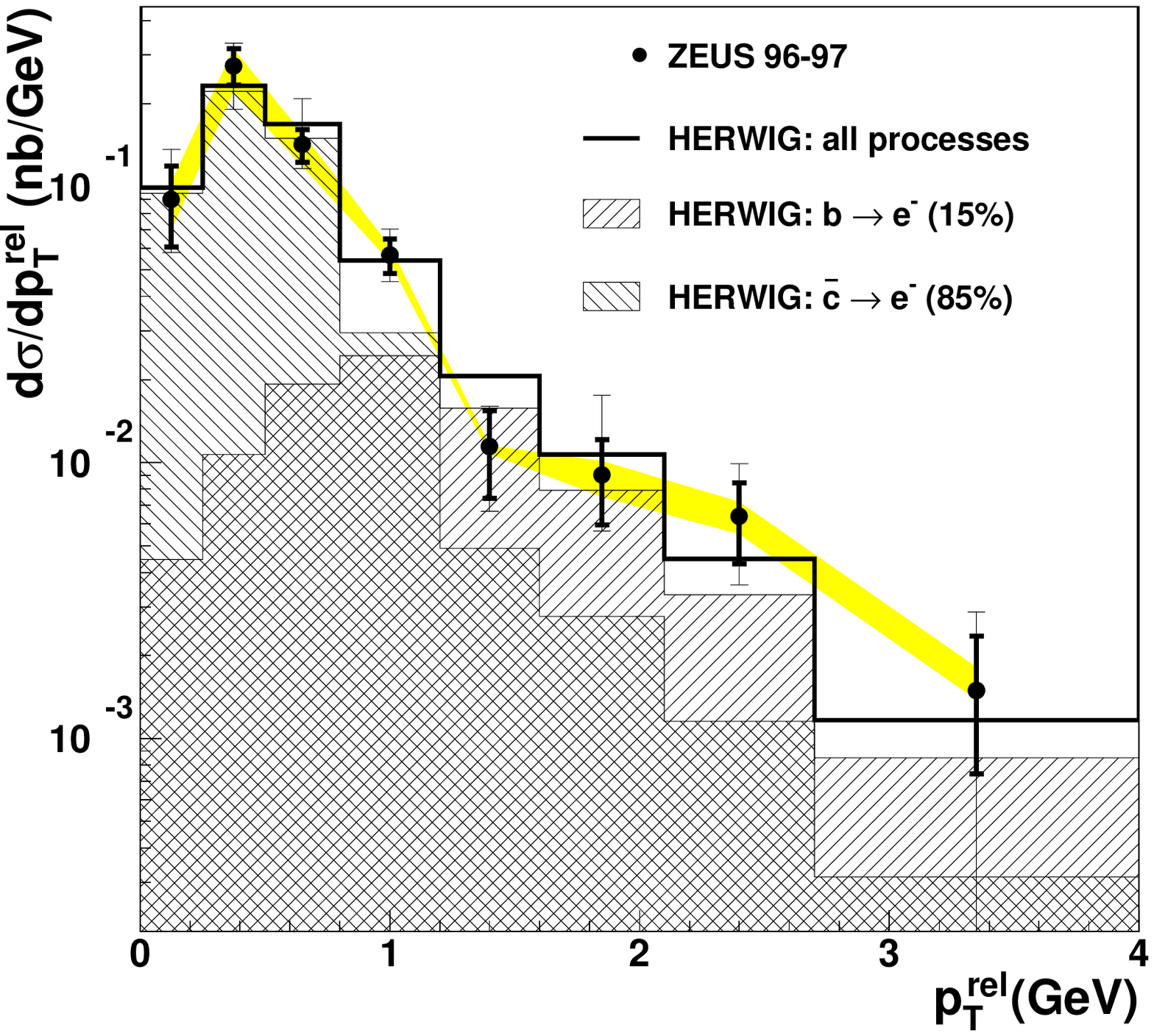} 
\end{minipage}
\begin{minipage}{5cm} 
\vspace*{0.3cm}
\epsfxsize=12.5pc 
\hspace*{0.5cm}\epsfbox{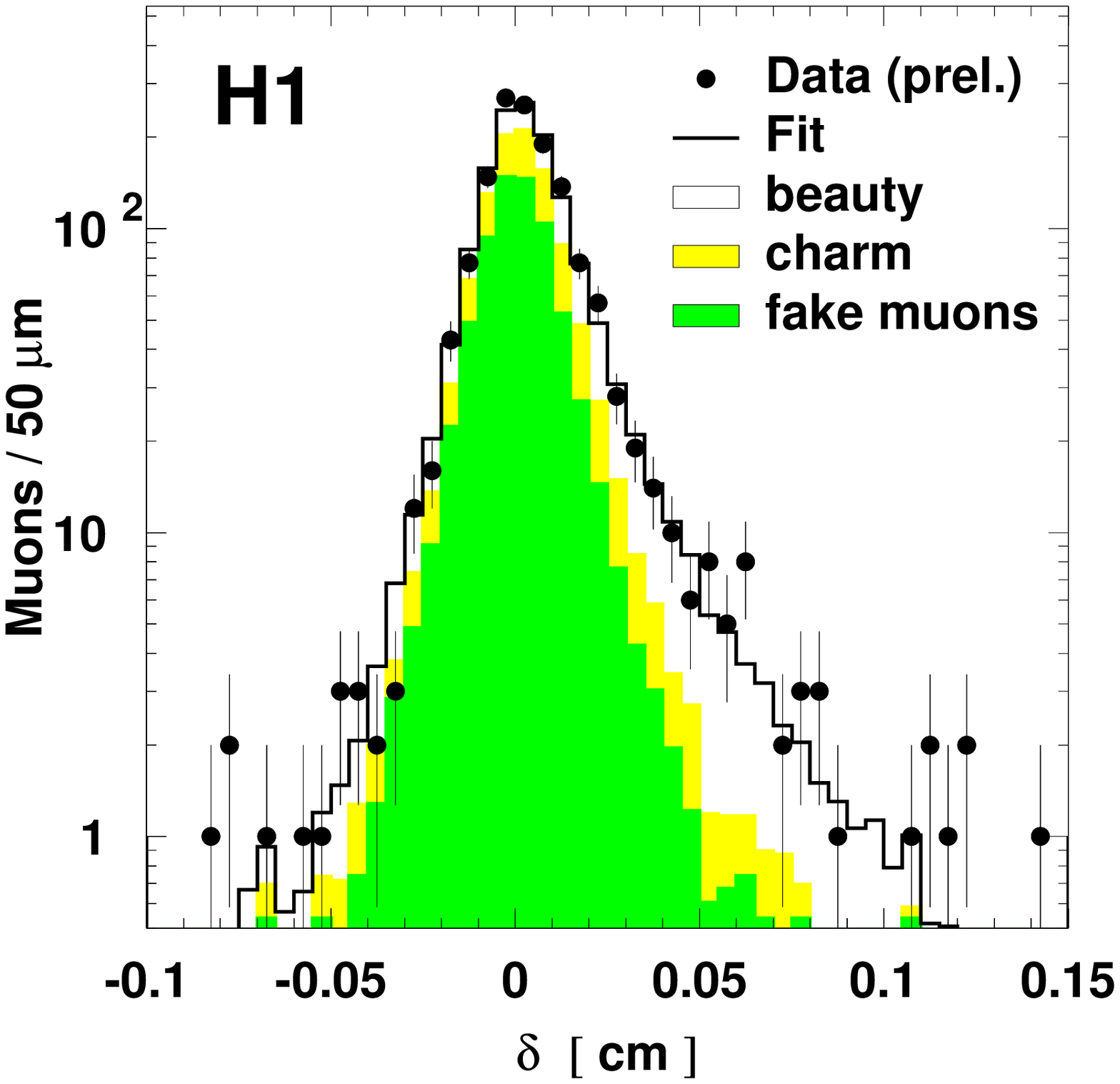} 
\end{minipage}

\setlength{\unitlength}{1cm}    
\begin{picture}(0.1,0.1)
\put(0.,0.6){(a)}
\put(6.2,0.6){(b)}
\end{picture}
\vspace*{-0.7cm}
\caption{(a)  
$d\sigma/dp_t^{rel}$ for ZEUS electron $\gamma p$ data and MC (normalised and fitted
to the data). 
(b)  Muon $\delta$ spectrum (H1 $\gamma p$ sample) with
the  decomposition from the $\delta$ fit. 
$\delta$ is positive if the track intersects the jet
downstream of the primary event vertex.
\label{fig:bgap}}
\end{figure}

\section{Open beauty production}
HERA results on open beauty production 
are based on 
semileptonic $b$ hadron decays within jets. The first
measurement\cite{h1openb}, obtained from an inclusive muon 
analysis on $\gamma p$ data recorded by H1 in 1996, yielded a  cross 
section exceeding the QCD expectation significantly. Since then, new 
data and improved experimental tools have become available. 

\subsection{Photoproduction}
ZEUS has  measured $b$ photoproduction in inclusive muon\cite{zeusmuon} 
and electron\cite{zeusopenb} analyses. 
As in the first  measurement at HERA,  
the transverse lepton momentum $p_t^{rel}$ relative to an associated jet 
is used as observable thus exploiting the large $b$ mass. 
Figure~\ref{fig:bgap}(a) shows the measured differential cross section, 
$d\sigma/dp_t^{rel}$ for a sample of $943$ electron candidates 
($p_{t}>1.6$ GeV, $|\eta| < 1.1$) from data collected during 1996 and 
1997 corresponding to an integrated luminosity of  
\mbox{${\cal L}=38.5$~pb$^{-1}$}. 
Using MC simulations to model the beauty signal and charm background, a 
fit  (taking the normalisation from the data) results in a beauty fraction 
$f_b$ of $15\%$ with the signal dominating the high $p_t^{rel}$ region.

\begin{figure}[b]
\vspace*{-0.9cm}
\epsfxsize=15.7pc
\begin{minipage}{6cm} 
\vspace*{0.2cm}
\hspace*{-0.4cm}\epsfbox{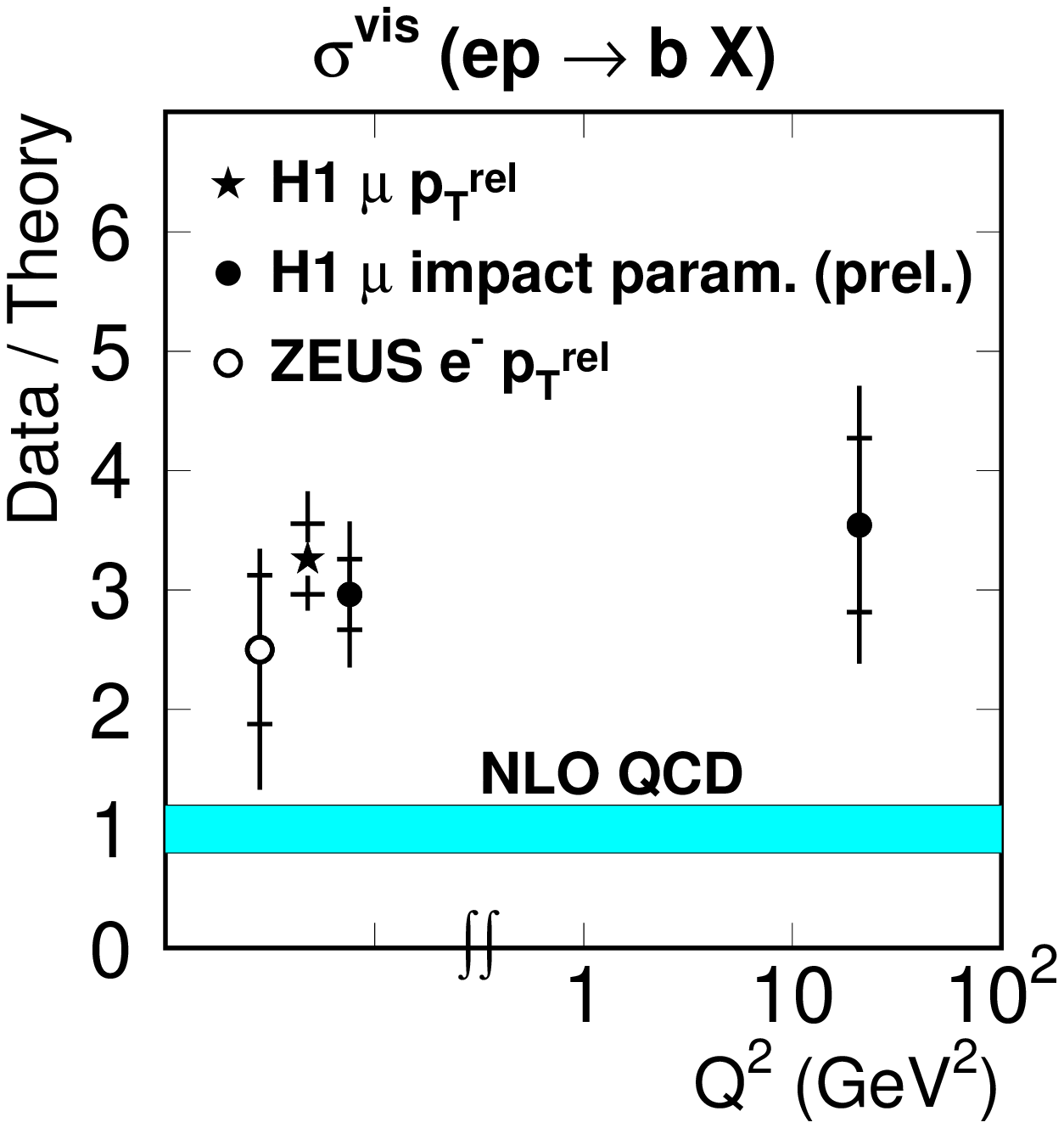} 
\end{minipage}
\begin{minipage}{5cm} 
\vspace*{0.0cm}
\epsfxsize=11.7pc 
\hspace*{0.6cm}\epsfbox{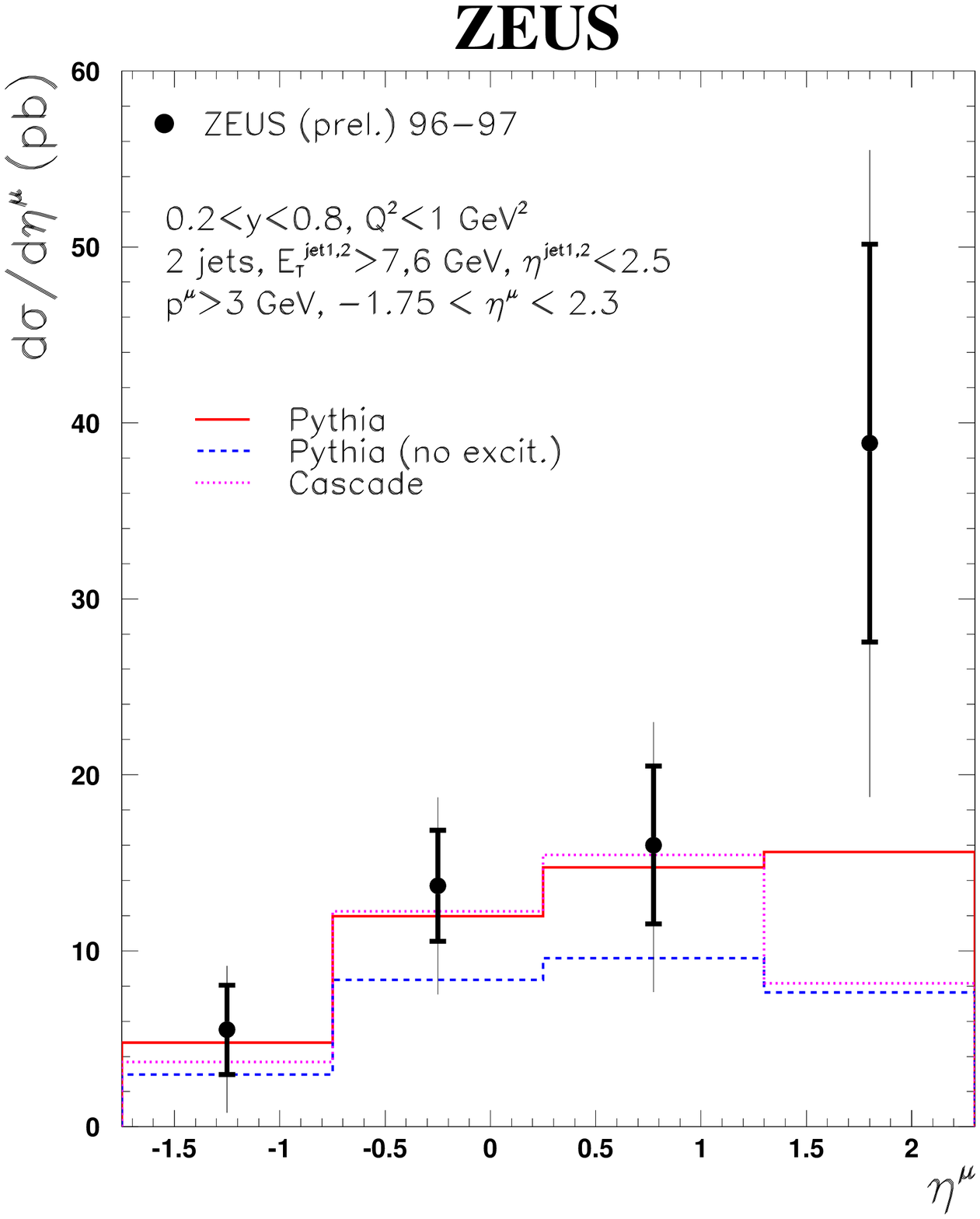} 
\end{minipage}

\setlength{\unitlength}{1cm}    
\begin{picture}(0.1,0.1)
\put(0.,0.6){(a)}
\put(6.2,0.6){(b)}
\end{picture}
\vspace*{-0.6cm}
\caption{(a) 
Ratio of measured $b$ production cross sections to theoretical predictions, as
a function of $Q^2$. 
(b) $b$ cross section as a function of $\eta(\mu)$, compared with LO MC predictions. 
\label{fig:bsum}}
\end{figure}
A recent H1 analysis\cite{h1bosaka} uses not only the $b$ mass signature but detects 
beauty hadrons also via their long lifetimes 
by reconstructing the signed impact parameter $\delta$.  
The $\delta$ distribution for  1415 muon candidates with
\mbox{$p_t>$ 2 GeV} and polar angles between $35^\circ$ and $130^\circ$ 
selected from 1997 data \mbox{(${\cal L}=14.7$~pb$^{-1}$)} 
is shown in figure~\ref{fig:bgap}(b) together with the result of a fit 
for the decomposition of the sample into contributions from beauty production, 
charm production and misidentified light hadrons ({\it fake muons}).
$f_b$ is determined to be $26\%$, the signal being enriched at high positive $\delta$. 
An improved measurement yielding a consistent result is obtained by 
combining both observables in a fit to the 
two--dimensional \mbox{($\delta$, $p_t^{rel}$)} distribution.

Translating $f_b$ into visible cross sections, all results are 
consistently above corresponding NLO QCD predictions obtained 
from the FMNR program,\cite{frixi} see figure \ref{fig:bsum}(a). 
The excess is significant for the H1 measurements.

In the search for the source of the discrepancies between data and theory 
valuable information might be gained from differential cross section
measurements. First results in the muon channel have recently been presented 
by ZEUS. 
Within errors,  a LO MC gives an acceptable description with a tendency 
to underestimate the cross section for large muon pseudorapidities, see figure  \ref{fig:bsum}(b).
Clearly, more statistics is needed to draw any firm conclusions.

\subsection{DIS} 
Having established the impact parameter method in the $\gamma p$ regime, 
H1 performs a combined \mbox{($\delta$,$p_t^{rel}$)} analysis\cite{h1bmoriond} 
to measure for the first time 
open $b$ production in DIS. Based on 171 muon candidates selected from 1997 data 
(${\cal L}=10.5$~pb$^{-1}$), the cross section in the visible range, defined by 
\mbox{$p_t(\mu)>2$ GeV}, $35^\circ<\theta(\mu)<130^\circ$, 
 $Q^2<1$ GeV$^2$ and $0.1<y<0.8\,$, is determined to 
$[39\;\pm\;8\;(stat.)\;\pm 10\;(syst.)\;]\;{\rm  pb}$.
This again exceeds significantly the NLO QCD prediction 
of $(11 \pm 2)\;{\rm pb}$,  which is  
obtained from the HVQDIS program,\cite{hvqdis} cf. figure \ref{fig:bsum}(a).

\section{Summary}
The study of open charm production at HERA is rapidly 
becoming a precision tool to test QCD in the presence of two hard scales. 
In the overall picture, pQCD gives a reasonable 
description of the data thus confirming the concept of charm 
production as a dominantly hard, gluon--induced process accounting 
for a significant part of the proton structure. 
Differential cross section  measurements  using several  channels and including 
jet--based observables provide a starting point 
to investigate the details of the production mechanism.

Open beauty production is measured in inclusive lepton
analyses including the  $b$ lifetime signature. Previous photoproduction results are 
confirmed and  improved, and the DIS cross section is measured for the first time.
All measured cross sections are found to be above the NLO QCD predictions, 
similar to the observations in $p\bar{p}$\cite{hadronb} and  $\gamma\gamma$\cite{ggb}
interactions.

\end{document}